\documentclass[prb,twocolumn,showpacs,showkeys]{revtex4} 
\usepackage{epsfig}
\newcommand{\cmt}{Hg$_{\scriptstyle 0.3}$Cd$_{\scriptstyle 0.7}$Te\,}

\newcommand{\czt}{Cd$_{\scriptstyle 0.96}$Zn$_{\scriptstyle 0.04}$Te\,}
\newcommand{\grad}{$^{\circ}$C\,}
\DeclareMathAlphabet{\mathitb}{OT1}{cmr}{bx}{sl}
\begin{document}

\title{Quasi-ballistic transport in HgTe quantum-well nanostructures}

\date{\today}
\author{V.~Daumer}
\email{daumer@physik.uni-wuerzburg.de}
\author{I.~Golombek}
\author{M.~Gbordzoe}
\author{E.~G.~Novik}
\author{V.~Hock}
\author{C.~R.~Becker}
\author{H.~Buhmann}
\author{L.~W.~Molenkamp}
\affiliation{Physikalisches Institut(EP 3), Universit\"{a}t
W\"{u}rzburg, Am Hubland, 97074 W\"{u}rzburg, Germany}

\begin{abstract}
The transport properties of micrometer scale structures fabricated
from high-mobility HgTe quantum-wells have been investigated. A
special photoresist and Ti masks were used, which allow for the
fabrication of devices with characteristic dimensions down to
$0.45~\mu$m. Evidence that the transport properties are dominated
by ballistic effects in these structures is presented. Monte Carlo
simulations of semi-classical electron trajectories show good
agreement with the experiment.
\end{abstract}


\maketitle

%\section{Introduction}
The giant Zeeman effect and Rashba spin orbit ({\sc s-o})
splitting\cite{Rashba} in magnetic two dimensional electron gas (2DEG)
structures have recently 
aroused much interest due to their possible application in
spintronics. These two effects are largest in the narrow gap II-VI
semiconducting materials based on HgTe with which high-mobility
quantum-well (QW) structures have been realized\cite{Zha01}. In
addition it is possible to replace Hg substitutionally by
isoelectronic magnetic atoms such as Mn, which enhances the
effective g-factor, i. e. $g^*\cong 50-60$, whereas the sample
mobilities are only weakly affected\cite{Zha03}. Therefore, HgTe
QW structures offer interesting opportunities to study spin
related transport effects. One goal is the exploration of the
electronic spin behavior in nanostructures in which transport is
dominated by  ballistic effects. However, up to now ballistic
transport has not been demonstrated in HgTe QW structures, mainly
due to specific material properties that prevent the application
of well known and established nano-structuring technologies used
for Si and GaAs based structures. 

Here we report the
observations of ballistic transport in HgTe. In a cross shaped
geometry with channel widths of 0.45 and 1.0~$\mu$m a non-local
resistance signal is detected. The signal can be explained
qualitatively and quantitatively by an approach based on the
Landauer-B\"uttiker~(LB) formula\cite{lb} and classical Monte
Carlo (MC) simulations which consider the sample geometry only.
The scattering time related to the transport mean free path is
used as a fit parameter. The scattering time $\tau$, which is
obtained from these simulations agrees well with that obtained
from transport measurement on macroscopic Hall bar structures.

%\section{Experimental Details}
The {\it n}-type asymmetrically modulation doped QWs were
epitaxially grown in a Riber 2300 MBE system on  \czt (001)
substrates\cite{Zha01,Gosch98}. After an approximately 60~nm thick
CdTe buffer layer, grown at 315~\grad, the CdI$_2$ modulation
doped HgTe QWs were grown at 180~\grad. The barriers consist of
\cmt layers.  The 9~nm thick iodine doped layer in the lower
barrier is separated from the quantum well by a 11~nm spacer. The
QW width is 12~nm. Sample parameters such as carrier
concentration, mobility and Rashba {\sc s-o} splitting were
obtained from ac and dc measurements of a standard Hall bar in an
$^4$He bath cryostat with magnetic fields perpendicular to the
2DEG  up to 7~T. The Hall bar pattern was defined by optical
lithography and wet chemical etching. A carrier concentration of
$\rm 1.7\times 10^{16}m^{-2}$ and a carrier mobility of $\rm
6.2~m^2/(Vs)$, which corresponds to a Fermi wave vector $k_F$ of
$3.27\times10^8{\rm m}^{-1}$ and a transport mean free path
$l_{mfp}$ of $1.3\times 10^{-6}{\rm m}$, were obtained for the
sample discussed here. Concluding from these results, it should be
possible to observe ballistic transport effects in devices
fabricated from this material with an active area of less than
1~$\rm \mu m^2$. 

In standard nano-structuring lithographical
processes, polymethylmethacrylate (PMMA) in conjunction with
electron beam lithography is used to fabricate such structures.
However, for epitaxially grown HgTe samples this is not possible
due to the high bake-out temperature necessary for PMMA of about
200\grad. Temperatures exceeding 100\grad cause deterioration of
the HgTe QW structures by interdiffusion of well and barrier
materials. As an alternative, we have used the photoresist
ARU~4060/3 (Allresist). This resist can be used not only for
optical but also for electron beam lithographical pattern
transfer. The advantage of this resist is the low bake-out
temperature. For our samples a bake-out of 2~min at 80\grad was
sufficient, which ensures that the sample structures remain
unaffected. 

Samples were fabricated which exhibit  a cross-shape
geometry with lead widths of $\sim$ 1.0~$\mu$m (structure~A) and
$\sim$ 0.45~$\mu$m (structure~B). A
%SEM\footnote{scanning electron microscope} photograph
scanning electron microscope image of structure~B is shown
in Fig.~\ref{q1819sem}. These crosses have been written using an
acceleration voltage of 2.5~kV. The positive resist was developed
and Ti was evaporated onto the sample to serve as an etch mask for
the subsequent wet chemical etching process. Ti masks must be used
to avoid the strong under-etching that occurs with the use of a
simple photoresist mask together with the etchant described below.
Contact pads were fabricated in an optical lithography step with
standard optical photoresist (Microresist ma-P215). Both optical
and e-beam pattern were etched in a dilute solution of Br$_2$ in
ethylene glycol at room temperature for 30~s.
%In this process the sample was etched
%about 150~nm deep into the CdTe buffer layer.
After etching about 150~nm into the CdTe buffer layer, the
resist and the Ti mask were removed with aceton and a 2:1 H$_2$O:HF(50\%)
solution for 10~s, respectively. Ohmic contacts were fabricated by thermal
bonding with indium.
%The samples were mounted in the cryostat described above.

Quasi-dc, low frequency (13~Hz) ac measurements with an excitation
voltage of 150~$\mu$V were carried out using lock-in techniques. Various
contact combinations have been used to characterize the sample after the
etching process. In the Hall geometry (I:~1$\rightarrow$3, V:~2$\rightarrow$4,
c.f. Fig.~\ref{q1819sem}) the carrier concentration was found to be the same
as that of macroscopic samples and therefore shows clearly, that the sample
properties have not been changed by the fabrication process.

In order to demonstrate that the transport properties are
dominated by ballistic effects we have performed non-local
transport measurements in different contact arrangements, which
previously have been demonstrated in high mobility
GaAs nano-structures\cite{lwm90,Hei98}. One of the most prominent
effects is the non-local {\it bend} resistance (NLR). This
signal is measured by passing current through contacts 1 and 2,
while the voltage is measured between contacts 3 and 4 (see
Fig.~\ref{q1819sem}). The bend resistance is obtained simply by
dividing the voltage V$_{3,4}$ by the injected current. If the
transport were dominated by diffusive scattering, no voltage
signal would be expected to appear between contacts 3 and 4 in
this geometry, whereas in the ballistic regime, electrons injected
from contact~1 into the cross reach the opposite channel before
they are scattered. This leads to charge accumulation at contact
area~3 and thus to the NLR signal. Applying a small magnetic field
perpendicular to the 2DEG plane deflects the ballistic electrons
and the voltage signal between 3 and 4 decreases. 

The results are
shown for a 1.0 and 0.45~$\mu$m cross in Fig.~\ref{q1819sdh4}. The
NLR signal is indeed observed, which is direct evidence of
ballistic transport in this device. As expected, the signal
exhibits a pronounced maximum around $\rm B=0$. With an applied
field the signal decreases, exhibiting a large dip with a negative
NLR signal before it approaches zero in the high field range ($\rm
B>2$T). This behavior of the NLR signal can be qualitatively
understood by applying the LB formalism. In our geometry the
resulting NLR is derived to be as follows:
\begin{equation}
  \label{lb4}
(R_{12,34} = ) \frac{V_c}{I_i}=\frac{h}{2e^2}\frac{T^2-t_rt_l}{(t_r+t_l)(2T^2+2(t_r+t_l)T+t^2_r+t^2_l)}
\end{equation}
where in our notation, $T$ is the transmission probability of
electrons from contact 1 in contact 3, and  $t_l$ and $t_r$ are
the transmission probabilities from contact 1 to contacts 2 and 4
respectivly. 

Comparing this result with the data presented in
Fig.~\ref{q1819sdh4}, one can see that at zero  magnetic field the
NLR signal is dominated by electrons that travel ballistically
from contact 1 to 3 ($T^2$, Eq.(\ref{lb4})). This signal is
reduced by electrons which reach either the left or right contact
($t_rt_l$, Eq.(\ref{lb4})). At zero magnetic field this
corresponds to electrons  that are either injected outside the
acceptance angle of contact~3 or are scattered by unintentional
impurities. In a magnetic field the electrons are deflected due to
the Lorentz force either toward the left or the right contact,
which implies that the NLR should decrease and approach zero,
i.e., $T=0$ and, either $t_l=0$ or $t_r=0$. However, due to the
boundary scattering processes mentioned above, an intermediate
field regime exists where the signal becomes negative. In this
regime the so called rebound trajectories\cite{Hei98} may cause
the product $t_l t_r$ to exceed $T^2$. Enlarging the B field
further will {\it guide} all electrons to only one contact  ($T^2
\rightarrow 0$ and, either $t_r \rightarrow 0$ or $t_l$ \
$\rightarrow 0$) and the NLR becomes zero. 

In the inset of
Fig.~\ref{q1819sdh4} one can see that in the regime where
$R_{12,34}$ is expected to approach zero, Shubnikov-de Haas
oscillations, which are not included in Eq.~\ref{lb4}, are
superimposed on the signal. However, the ratio of the absolute
magnitude of the positive signal at $\rm B=0$ and the largest
negative value is rather small compared to the published results
for high mobility GaAs structures\cite{lwm90,Hei98,timp}. The main
reason for this difference is the comparatively short mean free
path, which is of the order of the device dimensions in the
present case. Therefore, it is plausible that impurity scattering
in the cross area increases the transmission probability to
contacts 2 and 4, leading to a reduction in the NLR signal at zero
magnetic field. This effect is also observed, when the device size
is increased; the NLR signal for the 1.0~$\mu$m structure is much
smaller than that for the 0.45~$\mu$m structure, as shown in
Fig.~\ref{q1819sdh4}. In a first approximation the ratio of the
signal for these two structures can be used to estimate the carrier
mean free path in the cross area. The signal is proportional to
the number of electrons that reach contact 3 ballistically
($\propto \exp(-L/l_{mfp})$) reduced by those electrons that are
scattered into the contacts 2 and 4 ($\propto 1-
\exp(-L/l_{mfp})$). Evaluating the values deduced from
Fig.~\ref{q1819sdh4} a $l_{mfp} \approx 1.2$ $\mu$m is obtained
which is in good agreement with the average mean-free path for the
macroscopic sample.

In order to put these considerations on a more quantitative basis
we have used a MC simulation of the classical electron
trajectories in which electrons with an arbitrary velocity
distribution are injected through contact 1 into the cross. In
this model the electrons are then specularly reflected at the
sample boundaries. The electrons that reach the individual
contacts are counted. This number is proportional to the
corresponding transmission probability\cite{Hei98,been}.
Quantitatively, the experimental results can not be fully
explained by purely ballistic transport and boundary scattering.
From the measurements (Fig.~\ref{q1819sdh4}) one observes that the
signal exhibits additional fine structure which is not induced by
electronic noise. This fine structure is fully reproducible and
stable in time provided the sample is kept at low temperatures. We
identify the fine structure as universal conductance fluctuation
(UCF) electronic interference effects due to the random
distribution of scatterers within the cross area. Evidently, MC
simulations of classical electron trajectories are not appropriate
to simulate these interference effects. 

However, the overall line
shape and amplitude of the NLR signal can be reproduced by MC
calculations when additional random scattering is considered, i.e.\ the
ballistic propagation is altered randomly for electrons that dwell
longer than the scattering time $\tau$ in the cross
area. Fig.~\ref{q1819mc} shows the resulting NLR curve. For a
scattering time of $\tau = 1.1\times 10^{-12}$~s a good agreement
with the experimental data is obtained. This value implies a mean
free path of $\approx 0.9$~$\mu$m and agrees well with the $\tau =
1.6\times 10^{-12}$~s ($\Rightarrow l_{mfp} = 1.3$~$\mu$m),
obtained from the macroscopic transport measurements and the value
deduced from the peak height ($l_{mfp}= 1.2~\mu$m) discussed
above. The diffusive scattering is likely due to remote ionized
impurity scattering. These results demonstrate that for the given
device dimensions, electrons either reach the contacts
ballistically or are scattered randomly. This implies that
transport in these structures is in the transition regime between
ballistic and diffusive transport, which is usually referred to as
quasi-ballistic transport.

In conclusion, we have presented the evidence of
quasi-ballistic transport in high mobility HgTe QW nano-structures
which are fabricated with a technology that overcomes the specific
problems of Hg containing devices. Furthermore, a quantitative
analysis of the non-local resistance measurements revealed that
the actual HgTe QW nanostructure samples allow for a detailed
study of the transition from a diffusive to a local ballistic
transport regime.

%\section{Acknowledgments}
The financial support of the Deutsche Forschungsgemeinschaft (SFB 410), the
Alexander von Humboldt Stiftung, and the DARPA SPINS program is gratefully
acknowledged.

%\clearpage

\clearpage

\begin{figure}[htb]
\caption{Scanning electron microscope photograph of a cross with 0.45~$\mu$m wide leads.}
\label{q1819sem}
\end{figure}

\begin{figure}[htb]
\caption{Non-local resistance signal (NLR) for structures with
lead widths of 1.0~$\mu$m (structure A) and $0.45~\mu$m (structure
B). The inset shows the NLR of the smaller device for an extended
magnetic field range.} \label{q1819sdh4}
\end{figure}

\begin{figure}[htb]
\caption{Experimental data for a $0.45~\mu$m device together with
the Monte Carlo simulation result (smooth curve) for a scattering
time of $\tau = 1.1\times 10^{-12}$ s.} \label{q1819mc}
\end{figure}

\clearpage

\setcounter{figure}{0}

\begin{figure}[htb]
\epsfig{figure=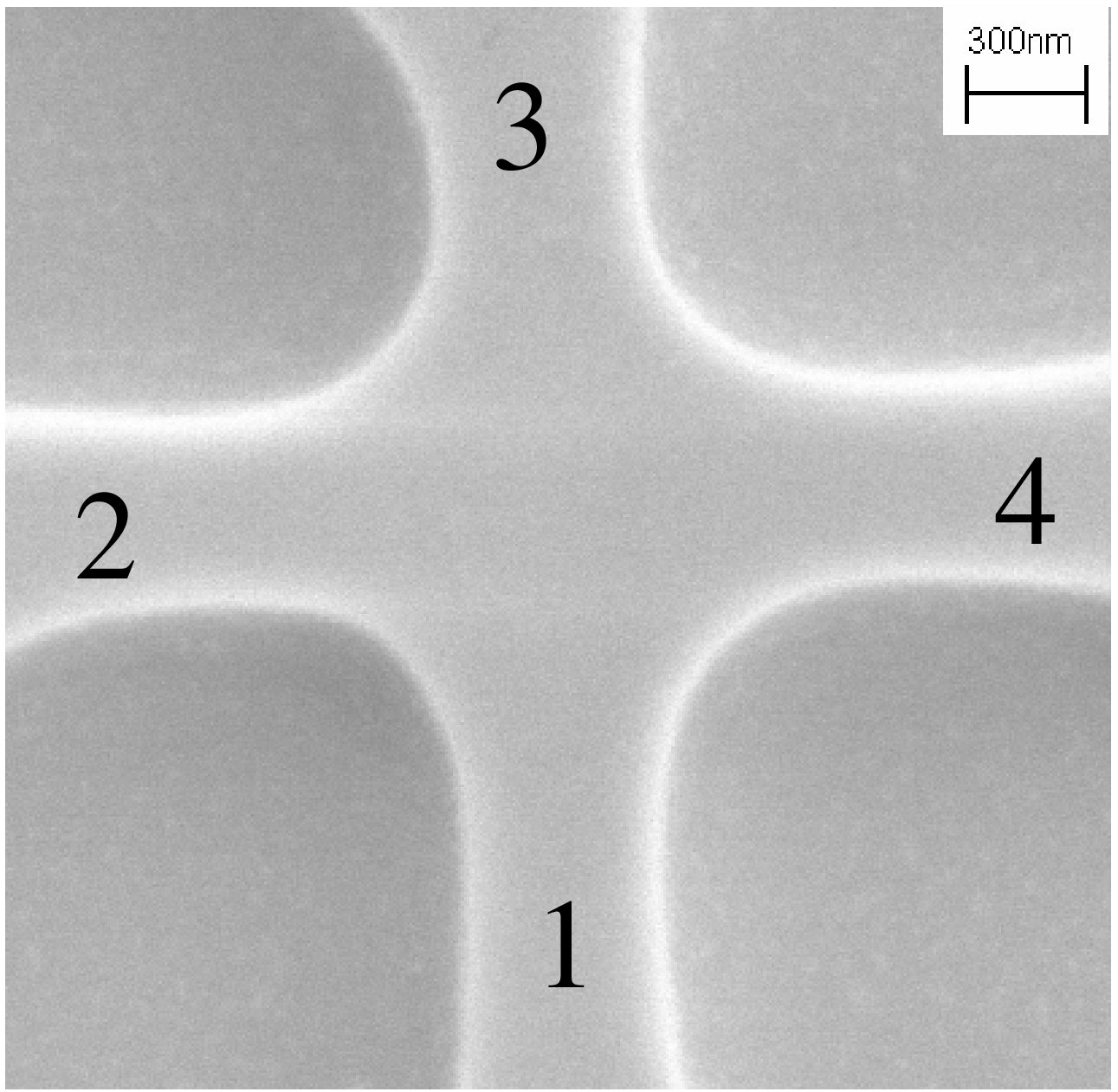,width=0.8\linewidth} 
%\caption{SEM photograph of a cross with 0.45~$\mu$m wide leads.}
%\label{q1819sem}
\end{figure}

\begin{figure}[htb]
\epsfig{figure=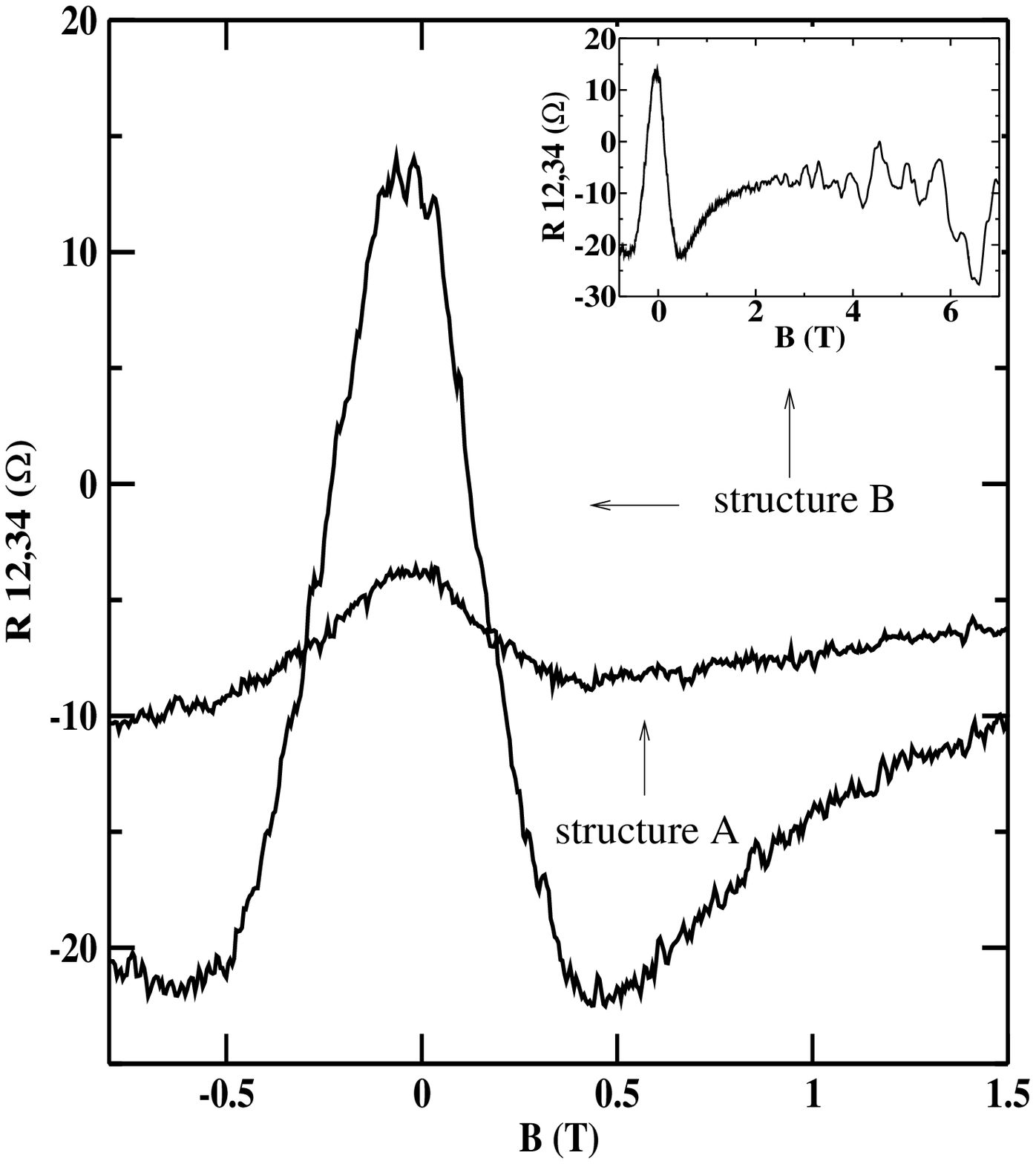,width=0.8\linewidth}
%\caption{Non-local resistance signal (NLR) for structures with lead widths of
%1.0~$\mu$m (structure A) and $0.45~\mu$m (structureB). The inset shows the
%NLR of the smaller device for an extended magnetic field range.}
%\label{q1819sdh4}
\end{figure}

\begin{figure}[htb]
\epsfig{figure=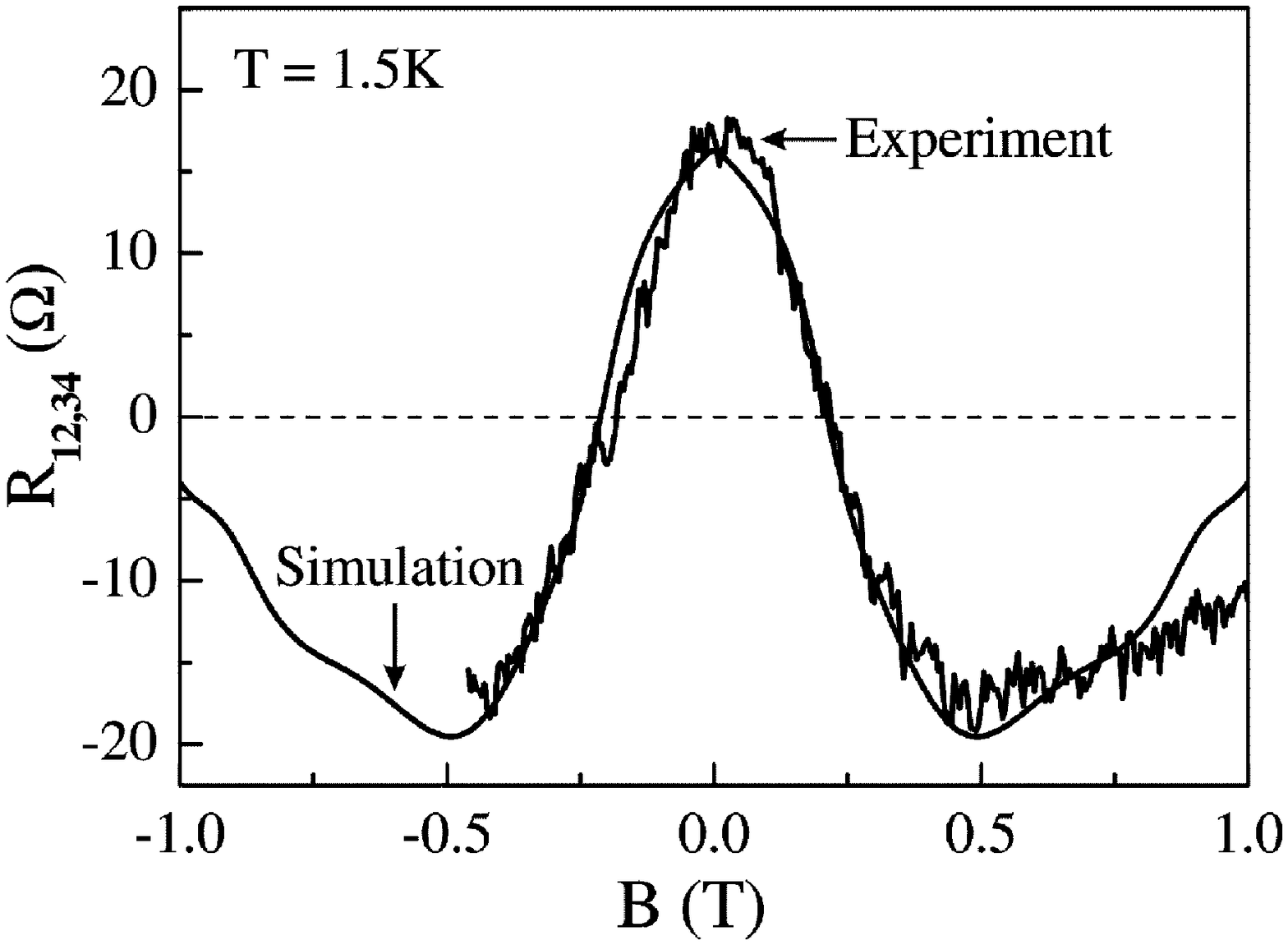,width=0.8\linewidth}
%\epsfig{figure=MonteCarlo.eps,width=0.8\linewidth}
%\caption{Experimental data for a $0.45~\mu$m device together with
%the Monte Carlo simulation result (smooth curve) for a scattering
%time of $\tau = 1.1\times 10^{-12}$ s.}
%\label{q1819mc}
\end{figure}

\end{document}